\documentclass[superscriptaddress, twocolumn, prl, nofootinbib]{revtex4-1}
\usepackage{amsmath, amssymb, color, stmaryrd, wasysym, esint}
\makeatletter
\newsavebox{\@brx}
\newcommand{\llangle}[1][]{\savebox{\@brx}{\(\m@th{#1\langle}\)}%
  \mathopen{\copy\@brx\kern-0.5\wd\@brx\usebox{\@brx}}}
\newcommand{\rrangle}[1][]{\savebox{\@brx}{\(\m@th{#1\rangle}\)}%
  \mathclose{\copy\@brx\kern-0.5\wd\@brx\usebox{\@brx}}}
\makeatother

\makeatletter
\newsavebox{\@brxx}
\newcommand{\lllangle}[1][]{\savebox{\@brxx}{\(\m@th{#1\langle}\)}%
  \mathopen{\copy\@brxx\kern-0.5\wd\@brxx\usebox{\@brxx}\kern-0.5\wd\@brxx\usebox{\@brxx}}}
\newcommand{\rrrangle}[1][]{\savebox{\@brxx}{\(\m@th{#1\rangle}\)}%
  \mathclose{\copy\@brxx\kern-0.5\wd\@brxx\usebox{\@brxx}\kern-0.5\wd\@brxx\usebox{\@brxx}}}
\makeatother

\usepackage{graphicx}
\usepackage{dcolumn}
\usepackage{bm}
\usepackage{xcolor}
\usepackage{xspace}
\usepackage[makeroom]{cancel}
\usepackage{float}
\usepackage{siunitx}
\usepackage{mathtools}
\usepackage{stackengine}

\definecolor{linkcolor}{rgb}{0,0,0.6} 
\usepackage[pdftex,colorlinks=true,
	pdfstartview = FitV,
	linkcolor    = linkcolor,
	citecolor    = linkcolor,
	urlcolor     = linkcolor,
	hyperindex   = true,
	hyperfigures = false]{hyperref}

\begin{document}

\title{Mesoscopic heterogeneity in biomolecular condensates from sequence patterning}

\author{Luke K. Davis}
\affiliation{%
Isaac Newton Institute, University of Cambridge, Cambridge CB3 0EH, UK
}%
\affiliation{%
Department of Mathematics, University College London, 25 Gordon Street, London WC1H 0AY, UK
}%
\affiliation{%
School of Mathematics and Maxwell Institute for Mathematical Sciences, University of Edinburgh, Edinburgh EH9 3FD, UK
}%

\author{Andrew J. Baldwin}
\affiliation{%
Physical and Theoretical Chemistry, Oxford University, South Parks Rd, Oxford OX1 3QZ, UK}%
\affiliation{Kavli Institute of Nanoscience Discovery, Dorothy Crowfoot Hodgkin Building, Sherrington Rd, Oxford OX1 3QU, UK}

\author{Philip Pearce}
\email{philip.pearce@ucl.ac.uk} 
\affiliation{%
Department of Mathematics, University College London, 25 Gordon Street, London WC1H 0AY, UK
}%



\begin{abstract}
\textbf{Abstract.} Biomolecular condensates composed of intrinsically disordered proteins (IDPs) are vital for proper cellular function, and their dysfunction is associated with diseases including neurodegeneration and cancer. Despite their biological importance, the precise physical mechanisms underlying condensate (dys)function are unclear, in part owing to the difficulties in understanding how biomolecular sequence patterns influence emergent condensate behaviours across relevant length and timescales. Here, through minimal physical modelling, we explain how IDP sequence patterning gives rise to nano-scale organisational heterogeneities in condensates. By applying our coarse-grained molecular-dynamics polymer model, which accounts for steric, attractive, and electrostatic interactions, we systematically quantify and map out the emergent morphological phases resulting from a wide range of sequence patterns. We demonstrate how sequences that enable local coil-to-globule transitions within regions of single polymers -- driven by a competition between the preferred crowding densities of different regions in the sequence -- also exhibit cluster formation in condensates. Overall, our work provides a conceptual framework to understand how sequence properties determine mesoscopic organisation within biomolecular condensates.
\end{abstract}

\maketitle

\textbf{Introduction--}Intrinsically disordered proteins (IDPs), though devoid of a stable tertiary structure, play vital roles in sub-cellular machinery \cite{Uversky2002, Wubben2020,Bondos2021,Alberts2022}, including in the formation of biomolecular condensates \cite{Banani2017,Brocca2020,Alberti2021,Choi2020}. Biomolecular condensates are micron-sized assemblies of proteins and macromolecules whose exterior is not encapsulated by a lipid membrane \cite{Alberti2021}. Present in eukaryotic and prokaryotic cells, condensates are a promising candidate to help explain how cells orchestrate complex reaction and kinetic pathways within a highly crowded and stochastic environment \cite{Banani2017}, and their dysregulation has been implicated in diseases such as neurodegeneration and cancer \cite{Spannl2019,Vendruscolo2022}. To better understand how condensates regulate biological functions in health and disease \cite{Mitrea2022}, we require a biophysical understanding of how multiple IDPs (and other biomolecules) are organised within these structures. 

Condensates are routinely characterised by optical microscopy methods with a resolution above the diffraction limit ($\sim$200~nm) \cite{Nott2015,Nott2016,Li2012,Guillen-Boixet2020} and their assembly thermodynamics is well explained by the continuum Flory-Huggins model \cite{Nott2015,Brangwynne2009}. Together, these experimental and theoretical observations have led to condensates being widely perceived as spatially uniform fluids; nevertheless, heterogeneities have been observed in condensates containing multiple macromolecular components \cite{Feric2016,Rana2024,Shrinivas2021}. More recently, oligomeric clusters of IDPs have been observed in the dilute phase \cite{Kar2022,Lan2023,Kar2024}. Moreover, there are indications of hierarchical nano- and meso-scale organisation inside condensates themselves comprising only a single protein component \cite{Erkamp2023, Dar2024}.

On a molecular basis, the contribution of individual residues to the thermodynamics of assembly have been characterised in detail for a limited number of phase-separating proteins, including Ddx4~\cite{Nott2015} and FUS~\cite{Bertrand2023,Wang2018}; however, we do not yet have a general method to reliably predict phase separation. Moreover, it is clear that the composition of residues alone does not dictate the tendency of a sequence to phase separate -- the distribution of residues within the protein is also important \cite{Nott2015,Martin2020,vonBulow2024}. Interactions between proteins have been investigated extensively using coarse-grained simulations~\cite{Lin2016,Lin2017,Dignon2018,Martin2020,Holehouse2021,Saar2021,Joseph2021,Mittag2022,Kar2022,Shillcock2022,Farag2023,SundaravadiveluDevarajan2024}, which collectively can be interpreted in terms of a  ``stickers and spacers" model \cite{Dill1985,Rubinstein1997,Harmon2017,Martin2020}. In such models, polymers are represented as beads on a chain, with regions that only interact through excluded volume (spacers), and regions that additionally interact through cohesion and charge (stickers). While there have been some observations of nano-scale heterogeneity in simulations~\cite{Martin2020,Holehouse2021,Rana2021,Kar2022}, we lack a model or physical intuition explaining why some sequences lead to condensates with nano- and meso-scale heterogeneity, and others do not.

To address this, here, we build a minimal polymer physics model that accounts for IDP sequence heterogeneity, and we perform coarse-grained molecular dynamics (MD) simulations. We uncover a rich phase-space of condensate morphologies whose mesoscopic organisation is predictably described by patterning of the polymer sequence. Our results reveal that a competition between preferred crowding densities of cohesive and spacer regions drive organisation at the polymer and condensate level.

\begin{figure}[t!]
\centering
\includegraphics[width=.99\linewidth]{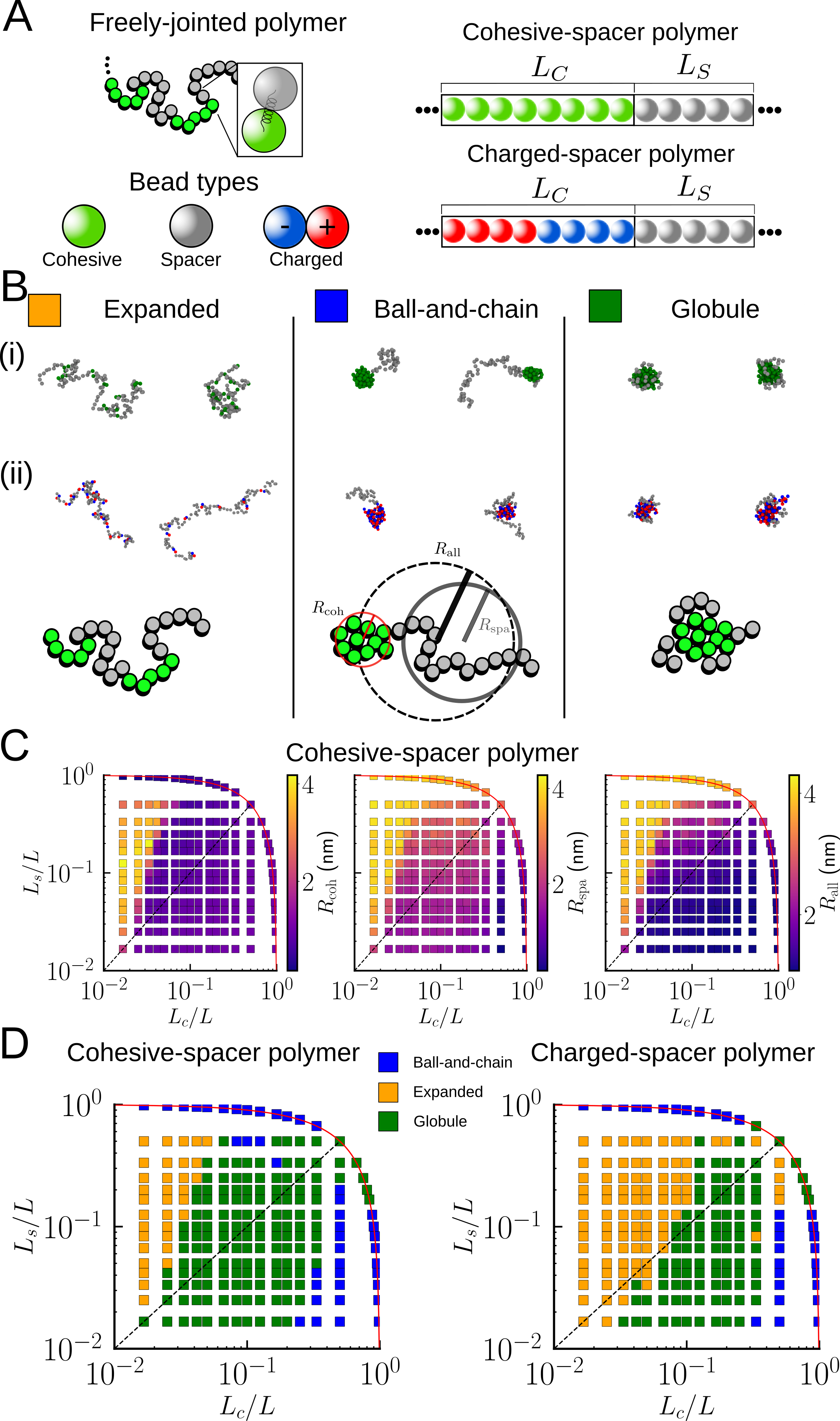}
\caption{\textbf{Coarse-grained IDP polymer model and single-polymer phase behaviour.} (\textbf{A}) Coarse-grained polymer model with monomer types for the polymers, which are connected by springs to form freely-jointed chains (left), and the different types of polymer repeating units, consisting of an attractive block of length $L_C$ and a spacer block of length $L_S$. (\textbf{B}) Qualitative classification of polymer morphologies for cohesive-spacer sequences (i) and charged-spacer sequences (ii). (\textbf{C}) Averaged radius of gyration, per polymer, of the cohesives $R_\text{coh}$, spacers $R_\text{spa}$, and all beads $R_\text{all}$ as a function of the sequence coordinates. (\textbf{D}) Quantitative single polymer phase diagrams for the cohesive-spacer (left) and charged-spacer (right) polymers as determined by inequalities (see main text).
}
\label{fig:Single}
\end{figure}

\textbf{Results--}To explore the key physical determinants governing the emergence of mesoscopic heterogeneity from sequence patterning in assemblies of disordered proteins (see Fig. \ref{fig:Single}A), we built a minimal coarse-grained molecular-dynamics polymer model consisting of freely-jointed beads that can be one of the following bead types: cohesive, charged, or spacer (full model details are given in the End Matter). 

We focused on periodic sequences where the repeating unit is one overall attractive block, which comprises either cohesive beads or regions of opposite charges, followed by one spacer block, which comprises only spacer beads. Under the assumption of either uncharged or charged units in the attractive block we may delineate any role that signed attractive molecular interactions play in determining condensate organisation. For example, it has recently been found that charge asymmetry controls the total size of condensates \cite{Luo2024}. For simplicity, we refer to the $L_C$-block, which can be made up of cohesives (green) or charges (blue and red), as the cohesive block. The lengths of a single cohesive block $L_C$ and that of a single spacer block $L_S$ are always kept fixed for a given sequence patterning. For a given $L_C$ and $L_S$ we also impose a fixed chain length $L$ where the repeating unit $L_C + L_S$ either divides exactly into $L$ or does not; in the latter case, the overhang is snipped off. Here we fix $L=120$, a highly composite number, so that we have a sufficiently high number (253) of exactly divisible and overhang sequences, built from all pairings of the divisors of $L$; the sequence is short enough to not be computationally cumbersome. Then, in our model, each sequence lives on a plane with unique coordinates $(L_C/L$, $L_S/L)$. A scalar describing each sequence is the symmetry measure defined as $\eta = (L_S-L_C)/L$ which describes the overall (im)balance between excluded-volume and cohesion in the sequence, and can be related to the sequence entropy \cite{Saar2021} (see Fig. S1). For a homopolymer (all cohesives) $\eta = -1$, for a perfectly balanced polymer $L_C = L_S$ and so $\eta = 0$, and for all spacers (only self-avoiding) $\eta = 1$.

We first examined how the sequence patterning of cohesive and spacer blocks of beads as defined in our polymer model (see Fig. \ref{fig:Single}A) governs the behaviour of single polymers ($N_p = 1$). We found that the resulting -- equilibrated -- single-polymer morphologies could be categorized into three distinct classes: (i) expanded: both the cohesive and spacer regions adopt more open conformations resulting in overall chain configurations that resemble an excluded-volume-dominated polymer \cite{RubinsteinColby2003}; (ii) ball-and-chain: cohesive beads form a well-defined spherical conformation and spacer beads form an expanded conformation resulting in an overall ``tadpole'' chain configuration \cite{Dyer2022}; (iii) globule: cohesive beads form a well-defined spherical conformation with a surrounding (heterogeneous) shell of spacer beads resulting in an overall compact polymer \cite{RubinsteinColby2003} (see Fig. \ref{fig:Single}B). To build a reproducible phase diagram, we first extracted the radius of gyration (size) of the spacer beads per polymer $R_\text{spa}$, of the cohesive $R_\text{coh}$ beads per polymer, and of all the beads per polymer $R_\text{all}$  (see Fig. \ref{fig:Single}C). We found that classification based on these quantities generated two phase diagrams -- one for cohesive-spacer polymers and one for charged-spacer polymers -- that reasonably matched with our qualitative observations: ``expanded'' if $\quad \frac{R_\text{coh}}{R_\text{all}} > \frac{1}{2}$, ``ball-and-chain'' if $\quad \frac{R_\text{coh}}{R_\text{spa}} < \frac{1}{4} \quad \text{or} \quad \frac{R_\text{spa}}{R_\text{all}} > \frac{3}{10}$, and ``globule'' otherwise (see Fig. \ref{fig:Single}D). In general, these results demonstrate the presence of multiple morphologies of single polymers determined by their patterning, in line with previous results \cite{Patel2008,Das2013}.

\begin{figure}[t!]
\centering
\includegraphics[width=.9\linewidth]{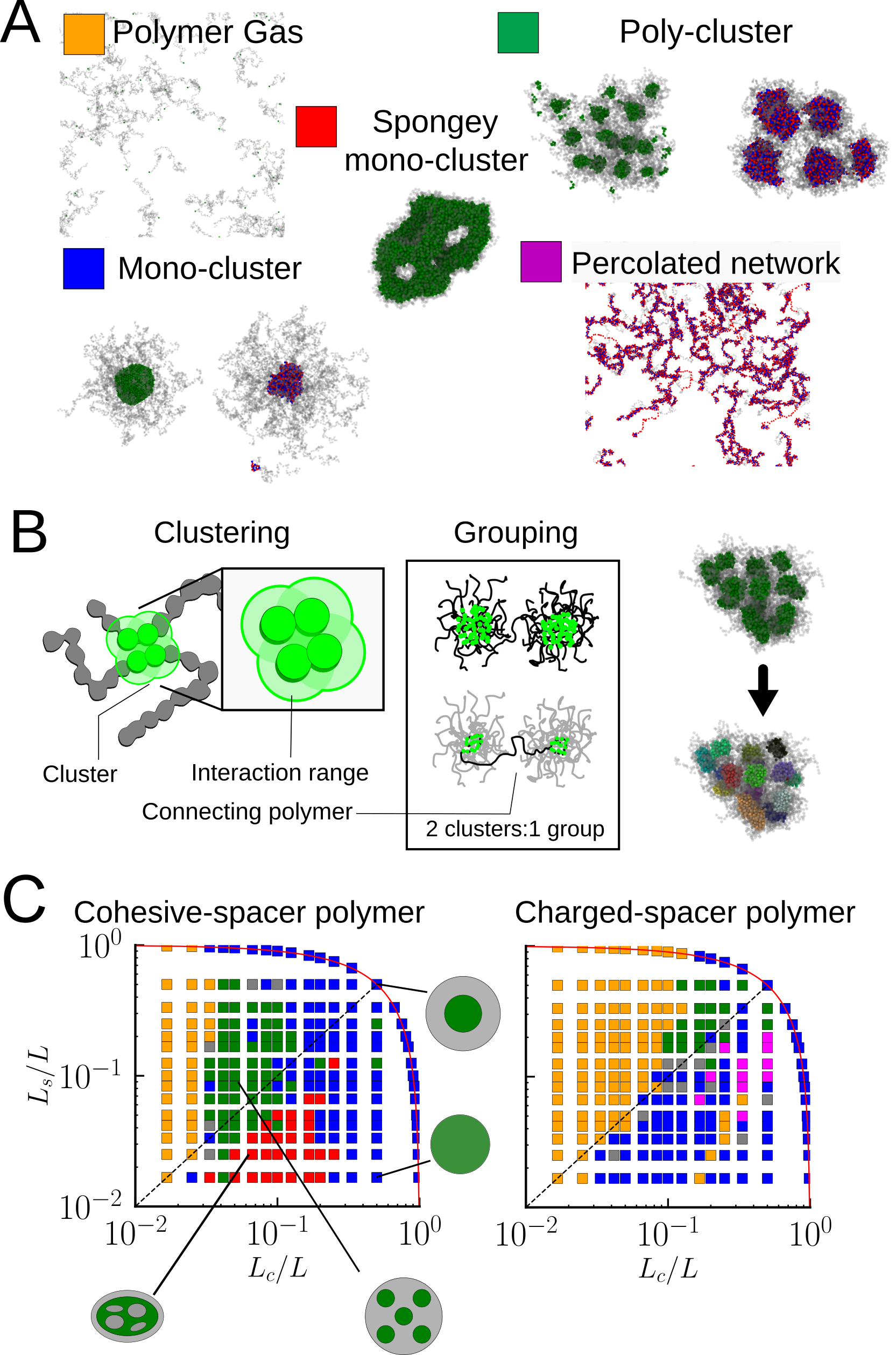}
\caption{ \textbf{Condensate morphologies and phase diagrams.} (\textbf{A}) Qualitative classification of collective morphologies resulting from visual inspection of snapshots of equilibrated many-polymer ($N_p=100$) simulations for selected cohesive-spacer polymer and charged-spacer polymer sequences. (\textbf{B}) (Left) Schematic diagram of the clustering and grouping procedure and (right) proof of concept on an example poly-cluster condensate. Colors label individual clusters. (\textbf{C}) Resulting quantitatively-determined phases for each of the cohesive-spacer polymer simulations (left) and charged-spacer polymer (right) simulations ($N_p = 100$).}
\label{fig:analysemany}
\end{figure}

To probe the effects of sequence patterning on modelled condensates we ran MD simulations of $N_p=100$ polymers, in the same conditions used for the single-polymer case. For some sequences, we found that cohesive or charged beads formed clear ``clusters'' within separate polymer ``groups'' (Fig. \ref{fig:analysemany}). Qualitatively, for each of the cohesive-spacer and charged-spacer polymers, we observed four different morphological regimes of which three are common (see Fig. \ref{fig:analysemany}A): (i) polymer gas: homogeneously distributed, hence not phase-separated, polymers that appear expanded as in Fig. \ref{fig:Single}B \cite{RubinsteinColby2003}; (ii)  mono-cluster: phase separation of polymers that form groups consisting of $\approx$ 1 cluster of cohesives, sometimes resembling reverse-micelles \cite{Floriano1999,RubinsteinColby2003,Safran2018}; (iii) poly-cluster phase separation of polymers that form groups each consisting of many clusters of cohesives \cite{Floriano1999,Panagiotopoulos2023}. For the fourth qualitative regime, the two polymer types are markedly different: (iv) spongey mono-cluster (cohesive-spacer only): phase separation of polymers that form groups of $\approx$ 1 cluster of cohesives with this cluster sometimes having holes and often being highly non-spherical -- the spacers tend to cover the cohesives \cite{Safran2018}; (iv) percolated network (charged-spacer only): accumulation of polymers into a fibrillated aggregate, which tends to span the entire box, in which charged beads comprise the majority of the fibrils, interspaced with spacer regions -- often strands of a single charge type are observed. We note here that actual disordered proteins, such as FUS, have been found to form fibrillated structures as well as condensates \cite{Bertrand2023}. 

We next wondered whether, as with the single-polymer case, the qualitatively identified morphological regimes could be systematically mapped back to the sequence coordinates $(L_C/L,L_S/L)$. Using the same observables used to quantify the single-polymers, i.e. $R_\text{spa}, R_\text{coh}$ and $R_\text{all}$ per polymer, we found that the data resembled the single-polymer data in some aspects but did not result in clearly identifiable regions in sequence space (see Fig. S2). This is because the emergence of new structural features such as clusters of cohesives (or charges), \emph{e.g.,} in Fig. \ref{fig:analysemany}A, cannot be solely encoded in single-polymer size metrics.

\begin{figure}[ht!]
\centering
\includegraphics[width=.99\linewidth]{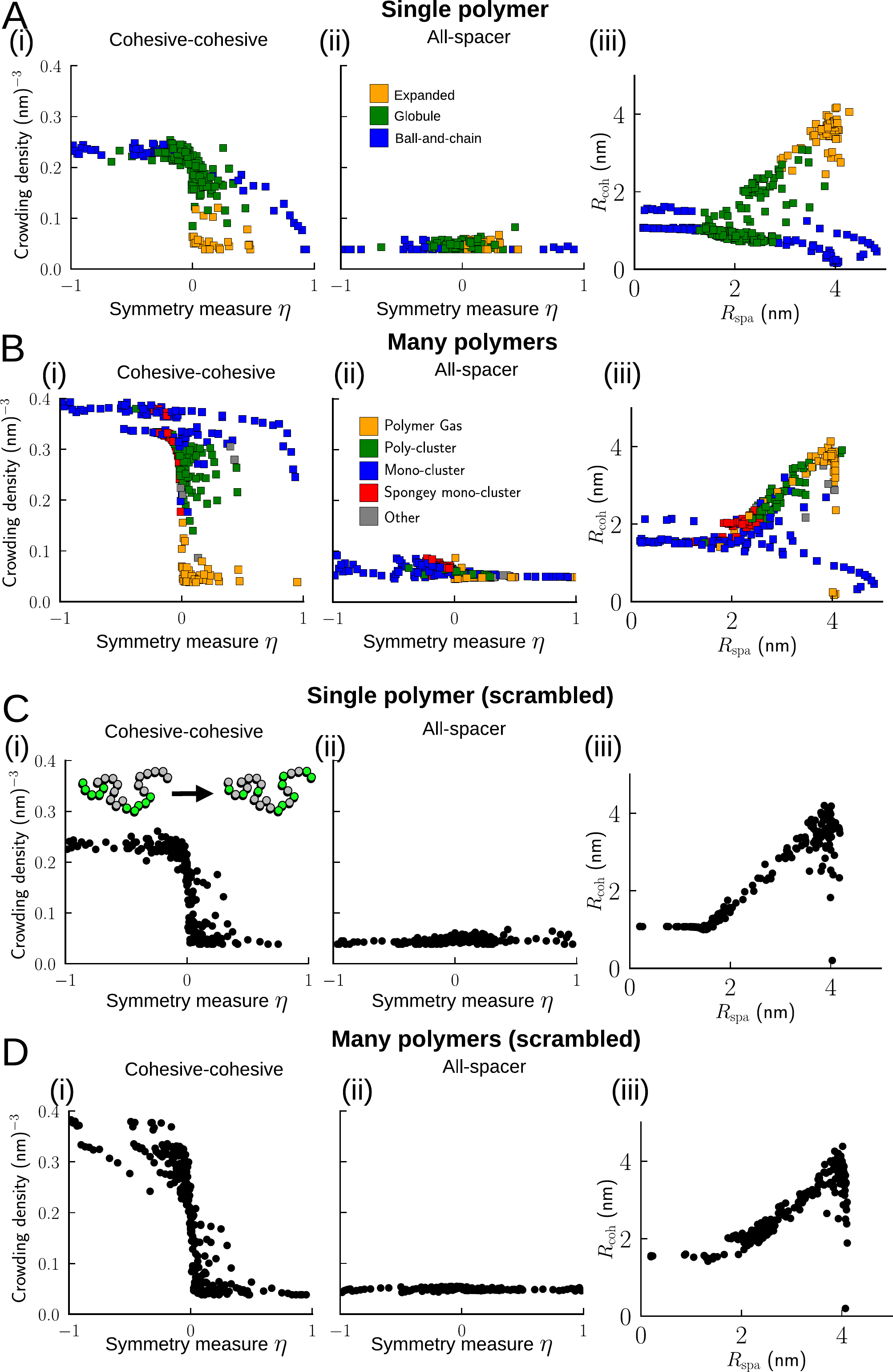}
\caption{\textbf{Crowding and local/global coil-to-globule transitions characterize polymer and condensate phase behaviours.} \textbf{(A)} Local crowding density as a function of the symmetry measure $\eta$ considering cohesive bead to cohesive bead (cohesive-cohesive) crowding (i) and spacer bead to all beads (all-spacer) crowding (ii), as determined from cohesive-spacer single-polymer simulations ($N_p=1$). (iii) Averaged radius of gyration of the cohesives $R_\text{coh}$ against that of the spacers $R_\text{spa}$. Colors indicate quantitatively determined phases as first shown in Fig.~\ref{fig:Single}C. (\textbf{B}) Same as (A) but for the many-polymer simulations ($N_p=100$); in panel (iii), $R_\text{coh}$ and $R_\text{spa}$ are per-polymer values. Colors indicate quantitatively determined phases as first shown in Fig.~\ref{fig:analysemany}C. (\textbf{C}) and (\textbf{D} are the same as (A) and (B) respectively, but for scrambled sequences.}
\label{fig:Final}
\end{figure}
Thus, to quantify the resulting organisation of the condensates we shifted our focus from individual polymer metrics to the characterisation of bead clusters and polymer groups (Fig. \ref{fig:analysemany}B). Briefly, for any instantaneous configuration of the polymers, clusters are found through hierarchical agglomeration of cohesive beads that are within the attractive, or oppositely-charged electrostatic (charges), interaction range, and groups are determined through heirarchical agglomeration of these clusters that have at least one connecting polymer (see Fig. \ref{fig:analysemany}B and section SI1 for details). The major advantages of this procedure are that it requires no specification of additional parameters, only the already specified interaction ranges, and that it is a deterministic rather than statistical algorithm.

Based on the quantification of the clusters and groups (see Fig. S3), we next sought to determine a reproducible phase diagram of the modeled condensate morphologies. To this end, for both the cohesive-spacer and charged-spacer polymers, we found several conditionals to classify the simulations into their respective four phases that agreed reasonably well with our qualitative observations (see SI2 for the inequalities and see Fig.~\ref{fig:analysemany}C for the phase diagrams). In Fig. \ref{fig:analysemany}C, there is a slight non-smoothness of classification in regions of parameter space where morphologies change, as is typical of configurations proximal to phase boundaries, and there is a strip of charge-based sequences classified in the polymer gas phase for a relatively large $L_c$, owing to those simulation configurations being almost indistinguishable from the polymer gas phase. It is not clear if these effects would still remain in the thermodynamic limit of infinitely many polymers. Overall, our simulations and analysis have yielded a reproducible phase diagram that relates sequence patterning properties (here $L_C$ and $L_S$) to emergent condensate organisation on the mesoscopic scale (Fig. \ref{fig:analysemany}).

 We next hypothesised that the observed collective morphologies arise from a competition between preferential cohesive-cohesive contacts and preferential higher available configuration space for the spacers. As a first test, we determined the local crowding densities of cohesives around other cohesives (cohesive-cohesive) and all types of beads around spacer beads (all-spacer) in single-polymer simulations of cohesive-spacer sequences (see Fig. \ref{fig:Final}A(i-ii)). Local crowding is determined via a voxel-based algorithm that essentially counts beads of a chosen type in the neighboring voxels of other beads of another chosen type and divides by the total voxel volume. We also extracted the relationship between the overall size of cohesive and spacer regions ($R_\text{coh}$ and $R_\text{spa}$, respectively) for all sequences (see  Fig. \ref{fig:Final}A(iii)). As the symmetry measure $\eta$ is lowered, we find that many sequences lie along a global coil-to-globule transition from a spacer-dominant regime (with symmetry measure $\eta > 0$, low cohesive-cohesive crowding and an ``expanded'' morphology) to a cohesive-dominant regime (with $\eta \leq 0$, high cohesive-cohesive crowding and a ``globule'' morphology; see Fig. \ref{fig:Final}A(i)). However, some sequences in the spacer-dominant regime ($\eta \geq 0$) exhibit high cohesive-cohesive crowding -- these mainly correspond to ball-and-chain morphologies, as well as globule morphologies in some cases. This suggests that cohesive blocks in these sequences have undergone a local coil-to-globule transition, while strictly maintaining low all-spacer crowding (see Fig. \ref{fig:Final}A(ii)). Indeed, for such sequences we find that the cohesive regions remain compact despite a large size of spacer regions (Fig. \ref{fig:Final}A(iii)). By contrast, for sequences that follow the global coil-to-globule transition, $R_\text{coh}$ and $R_\text{spa}$ are highly correlated (Fig. \ref{fig:Final}A(iii)). Overall, these results suggest that certain sequence patterns enable single polymers to undergo local coil-to-globule transitions that determine morphology. 

We next wondered whether the crowding behaviour in the single-polymer case ($N_p=1$) carried over to condensates with $N_p=100$ (see Figs.~\ref{fig:Final}B(i-iii)). Similarly to the single-polymer case, as $\eta$ is varied for condensates we find that many sequences lie along a clear transition from a spacer-dominant regime with low cohesive-cohesive crowding (mainly the ``polymer gas'' morphology) to a cohesive-dominant regime with high cohesive-cohesive crowding (various condensate morphologies) -- see Fig.~\ref{fig:Final}B(i). As before, we again find some sequences with high cohesive-cohesive crowding even in the spacer-dominant regime ($\eta>0$) -- these mainly correspond to mono-cluster and poly-cluster morphologies. All-spacer crowding and the radii of gyration are also found to be  qualitatively similar to the single polymer case (Figs.~\ref{fig:Final}B(ii-iii)). Furthermore, in terms of conditional probabilities, we are able to link the many-polymer phases to the single-polymer phases; specifically the polymer gas phase with the expanded phase, the poly-cluster phase with the globule phase, and the mono-cluster phase with the ball-and-chain phase (see Fig. S4). This suggests that local coil-to-globule transitions in single polymers appear as local heterogeneities (or clusters) in multi-polymer condensates.

\begin{figure}[t!]
\centering
\includegraphics[width=.99\linewidth]{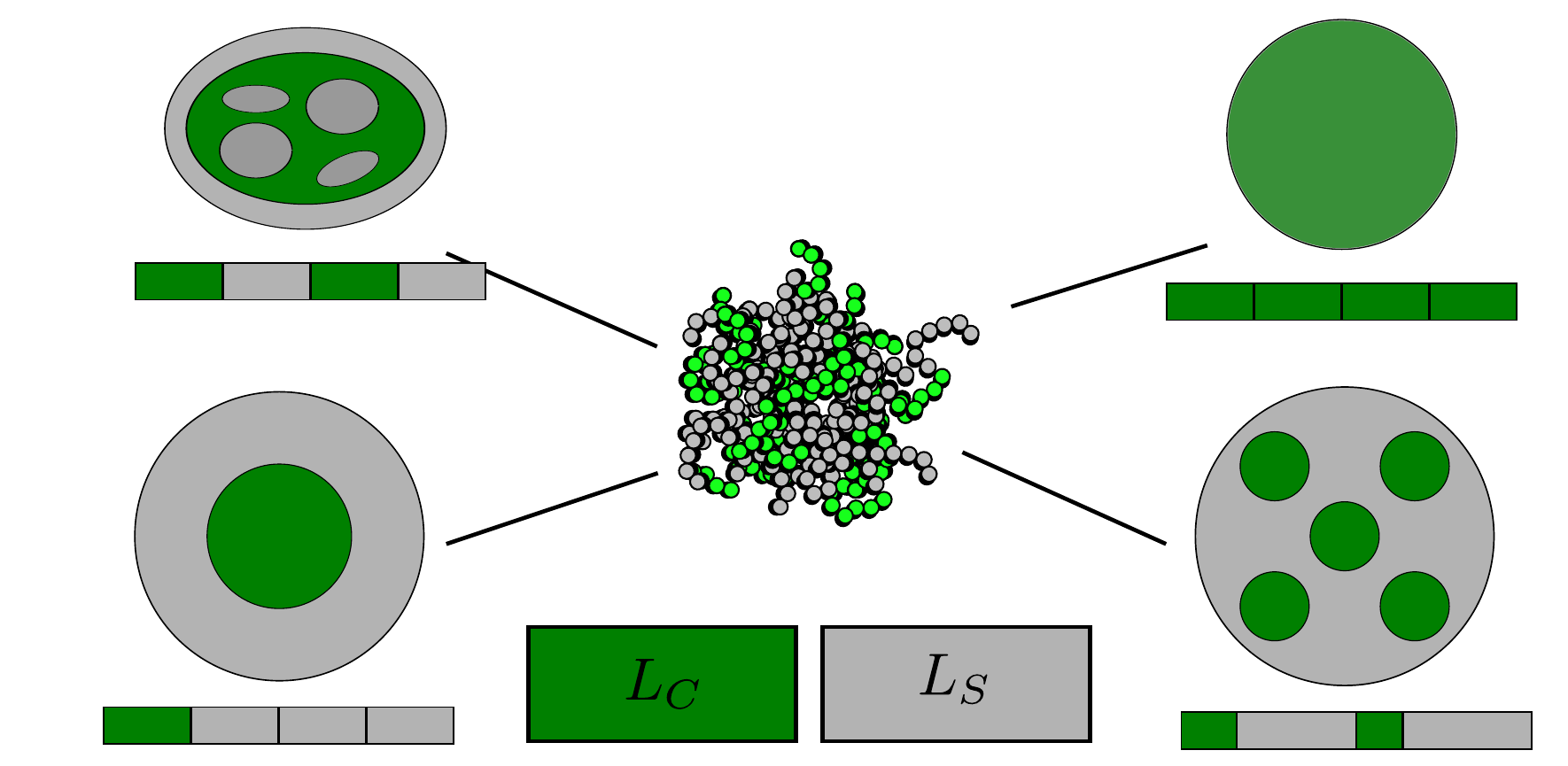}
\caption{Different condensate internal organization is favourable, for a given sequence patterning, if it minimizes the potential energy (or enthalpy) by enhancing cohesive-cohesive crowding whilst maximizing entropy, \emph{i.e.,} minimizing all-spacer crowding to maximize the number of available configurations of the spacer regions.}
\label{fig:summary}
\end{figure}

To further understand how sequence patterning determines polymer and condensate morphologies, we performed simulations of randomly scrambled sequences (at fixed $N_\text{coh}/N_p$ and $N_\text{spa}/N_p$); see Figs. \ref{fig:Final}C-D. Broadly, we found that scrambling increases the overall polymer radius of gyration (see Fig. S5), in line with previous work \cite{Bowman2020}, and is consistent with very low all-spacer local crowding (see Figs. \ref{fig:Final}C(ii),D(ii)). Furthermore, we found that all scrambled polymers lie broadly along the global coil-to-globule transition curve, controlled by the symmetry measure $\eta$, i.e. the overall amount of cohesion. In particular, $R_\text{coh}$ and $R_\text{spa}$ are highly correlated (Fig. \ref{fig:Final}C); there is a regime where $R_\text{coh}$ remains close to zero, which corresponds to $L_C \rightarrow 0$. We also see a similar effect on condensate morphologies (Fig. \ref{fig:Final}D). These results suggest that scrambling effectively homogenizes the polymer, precluding both local coil-to-globule transitions and well-defined local clusters in condensates.

\textbf{Discussion--}In this work, we use minimal polymer physics modeling to reveal the emergence of heterogeneous structures at multiple scales -- certain sequences form cohesive (or charged) ``clusters'' within single polymers and condensates. We find that preferential crowding of cohesives (or charges) and preferential spreading-out of spacers is a major driving force for the observed organisation. In particular, certain sequence patterns enable regions of cohesive beads to undergo a localized, as opposed to global \cite{Lin2017,dignon2018relation}, coil-to-globule transition whilst maintaining low spacer crowding. The links that we draw here between local coil-to-globule transitions in single polymers and cohesive clustering in mesoscopic condensates align with previous work, in which global coil-to-globule transitions in polymers have been found to predict overall condensate phase behaviour \cite{Nott2015,Lin2017,dignon2018relation}. Our finding that longer cohesive blocks in the sequence promote condensate heterogeneity complements previous related observations \cite{Martin2020,Holehouse2021,Rana2021,Kar2022} by clearly identifying the polymer- and condensate-scale processes that determine mesoscopic organisation. Furthermore, our results relating local crowding to polymer and condensate organisation can be explained in thermodynamic terms: higher cohesive contacts promote lower values of the internal energy, and lower densities around spacers promote higher values of the entropy (see Fig. \ref{fig:summary}). This enables patterned (as opposed to scrambled) polymers to form local clusters in both single polymers and mesoscopic condensates.

Our results have clear implications for emergent properties and processes in condensates including surface tension and viscosity \cite{Wang2021,Law2023}, aging \cite{Biswas2024}, thermodynamics \cite{An2024}, and chemical reactions \cite{Kirschbaum2021,Kilgore2022}. By establishing the polymer physics that connects sequence patterning to condensate organisation, this work contributes to the overarching goal of understanding how condensate biophysics, from the sequence to the emergent condensate scale, relate to biological function and dysfunction.

\textbf{Acknowledgements--}We thank Lindsay Baker, Emma Silvester, and Eugene Lin for helpful discussions. P.P. is supported by a UKRI Future Leaders Fellowship [MR/V022385/1]. L.K.D. acknowledges funding from the Isaac Newton Institute for Mathematical Sciences Postdoctoral Research Fellowship (EPSRC Grant Number EP/V521929/1). A.J.B. has received funding from the European Research Council (ERC) under the European Union’s Horizon 2020 research and innovation programme (grant agreement No 101002859). The authors acknowledge the use of the UCL Myriad High Performance Computing Facility (Myriad@UCL) and computing facilities at the UCL Department of Mathematics, and associated support services.

\textbf{Author contributions--}Conceptualisation: L.K.D, A.J.B., P.P.; Methodology: L.K.D, P.P; Data generation and analysis: L.K.D., Investigation: L.K.D, A.J.B, P.P.; Writing - Original Draft: L.K.D; Writing - Review \& Editing: L.K.D, A.J.B., P.P. The authors declare no competing interest.


\appendix
\onecolumngrid
\section{End Matter: Model and simulation details} 
\label{meth:model}
We model the IDP as a freely-jointed polymer (see Fig. \ref{fig:Single}A) consisting of $L=120$ bonded beads of diameter $\sigma = 0.4$~nm, which is the approximate diameter of an amino acid \cite{CarrionVazquez1999,Davis2020}, with the bonds being imposed through a harmonic spring potential given as
\begin{equation}
    U_b(r_{ij}) = \frac{k}{2} (r_{ij} - \sigma)^2, \quad |i-j| = 1,
    \label{eq:bonds}
\end{equation}
where $r_{ij}$ is the distance between particles $i$ and $j$, $k = 25$~$k_B T/\text{nm}^2$ is the spring constant with the value being chosen solely because it is large enough as compared with 1~$k_B T/\text{nm}{}^2$ (where $k_B$ is Boltzmann's constant and $T$ is the temperature).

Alongside the bonds, each bead experiences excluded-volume interactions with every other and may, depending on bead type, experience an attractive interaction with other beads. In the setup of our general model we will allow for the attractive interaction to arise either from a simple isotropic cohesive interaction (coh) or from opposite charges (cha). We include charged interactions separately to explore how effects between charged (signed) units in the polymer and/or repulsion (beyond excluded-volume) between like charges would result in different behaviour as compared with the simpler cohesive case. Beads that do not experience attraction, and therefore only excluded-volume interactions, are deemed spacers (spa). All of these -- independently tunable -- interactions are implemented through the following pair-potentials:

\begin{equation}
V_\text{spa}(r_{ij}) =
\begin{cases}
   U_\text{vol}(r_{ij}), \quad r_{ij} \leq r_\text{vol},\\
    0, \qquad  r_\text{vol} < r_{ij} ,\\
\end{cases}
 \label{eq:intermolecular1}
\end{equation}
and for the cohesives:
\begin{equation}
V_\text{coh}(r_{ij}) =
\begin{cases}
   U_\text{vol}(r_{ij}) + U_\text{coh}(r_{ij}), \quad r_{ij} \leq r_\text{vol},\\
U_\text{coh}(r_{ij}), \qquad r_\text{vol} < r_{ij} \leq r_\text{coh} ,\\
    0, \qquad  r_\text{coh} < r_{ij} ,\\
\end{cases}
 \label{eq:intermolecular2}
\end{equation}
and, finally, for the charged beads:
\begin{equation}
V_\text{cha}(r_{ij}) =
\begin{cases}
   U_\text{vol}(r_{ij}) + U_\text{cha}(r_{ij}), \quad r_{ij} \leq r_\text{vol},\\
U_\text{cha}(r_{ij}), \qquad r_\text{vol} < r_{ij} \leq r_\text{cha} ,\\
    0, \qquad  r_\text{cha} < r_{ij} ,\\
\end{cases}
 \label{eq:intermolecular3}
\end{equation}
where $r_\text{vol} = 2^{1/6} \sigma$ is the excluded-volume interaction range, $r_\text{coh} = 2.5 \sigma \equiv 1$~nm is the cohesive interaction range (for similar values used for coarse-grained models of IDPs see \cite{Davis2021,Joseph2021}), $r_\text{cha} = 7.5\sigma \equiv 3$~nm is the charged interaction range that is chosen due to the negligibly small energy values at this pair-wise distance ($U_\text{cha} \lessapprox 5\times10^{-4}$~$k_BT$, see also \cite{Joseph2021}). The functions $U_\text{vol}$, $U_\text{coh}$, and $U_\text{cha}$ are the respective excluded-volume, cohesion, and charged interaction pair-potentials, that are based on the Lennard-Jones potential, the truncated and shifted Lennard-Jones potential \cite{Allen2017}, and the screened-coulomb potential respectively. These functions are thus given as
\begin{align}
        U_\text{vol}(r) &= 4 \varepsilon_\text{vol} \left(\left( \frac{\sigma}{r} \right)^{12} - \left(\frac{\sigma}{r} \right)^6 \right) + \varepsilon_\text{vol}, \\
        u(r) &= 4 \varepsilon_\text{coh}  \bigg[ \left( \frac{\sigma}{r} \right)^{12} - \left(\frac{\sigma}{r} \right)^6 \bigg], \nonumber \\
        U_\text{coh}(r) &= \varepsilon' \left[ u(r) - u(r_\text{coh}) - (r-r_\text{coh})\left( \frac{d u(r)}{dr}\right)_{r=r_\text{coh}} \right], \\
        %
        %
        U_\text{cha}(r) &= \frac{e^2 q_i q_j}{4 \pi \varepsilon_0 \varepsilon_d r} \exp(-\kappa r),
        \label{eq:intermolecular4}
\end{align}
where $r=r_{ij}$, $\varepsilon_\text{vol}=50$~$k_B T$ is the strength of volume-exclusion, a value that is large enough (as compared with the thermal background $1~k_BT$) to impose a volume-exclusion diameter of $\sigma$ to within $\lessapprox 1\%$ and larger than the bond energy, $\varepsilon'$ is a conversion factor to ensure the truncated and shifted pair potential $U_\text{coh}$ has a minimum of $\varepsilon_\text{coh}$, $e=1.6021\times 10^{-19}$~\unit{\coulomb} is the charge of the electron, $\varepsilon_0\varepsilon_d=8.8542 \times 10^{-11}$~\unit{\farad\per\metre} is the rescaled (by the dielectric constant) permittivity of free space, $q_i = \pm 1$ is the unit charge of bead $i$, $C=4.11\times10^{-12}$ is a factor to convert from Joules to $k_B T$, and $\kappa^{-1}=2.5\sigma$~nm ($\equiv 1~\text{nm} \equiv 1000~\text{pm}$) is the Debye screening length, a commonly used value that corresponds to a monovalent salt concentration of $\sim0.1$~M \cite{Dignon2018,Joseph2021}. We note that the dielectric constant $\varepsilon_d=10$ simply scales the charged interactions with values ranging between $\sim$10-100 in the disordered protein modelling literature \cite{Dignon2018,Joseph2021}.

We sample configurations in the canonical ensemble ($NVT$) where a fixed $N_p$ polymers reside in a cube of fixed volume $V = l^3$ ($l=80$~nm), with periodic boundary conditions, and at a fixed temperature $T$. We time evolve the system according to overdamped Langevin dynamics given by
\begin{equation}
    \begin{aligned}
        \dot{\mathbf{r}}_i = -\mu \pmb{\nabla}_i U(\lbrace \mathbf{r} \rbrace) + \sqrt{2\mu k_B T} \pmb{\zeta}_i,
    \end{aligned}
    \label{eq:dyn}
\end{equation}
where $i = \lbrace 1, \ldots, N \rbrace$ ($N=N_p L$), $\mu$ is the mobility, and $\pmb{\zeta}_i$ is a white noise with zero mean and unit variance. 

We numerically integrate \eqref{eq:dyn} in the \texttt{LAMMPS} package \cite{Thompson2022} using $D = \mu k_B T = 10^{-9}$~\unit{\square\metre\per\second}, which is -- approximately -- the diffusion coefficient of a typical amino acid \cite{Ma2005}. Unless stated otherwise, for the production runs we use $\tilde{t} = t_\text{tot}/\delta t = 10^7$ integration steps, where $t_\text{tot}$ is the total time and $\delta t=0.004$ is the dimensionless timestep. To avoid long nucleation (for the condensates) times we initialize our simulations with polymers dragged to the origin via artificial forces on the cohesive (or charged) beads. After initialization, we perform $5\times10^6$ relaxation timesteps. We perform all simulations using dimensionless quantities.

\twocolumngrid
\bibliography{references-PR}

\end{document}